# Exceptional Points of Degeneracy Induced by Linear Time-Periodic Variation

Hamidreza Kazemi, Mohamed Y. Nada, Tarek Mealy, Ahmed F. Abdelshafy, and Filippo Capolino
*Department of Electrical Engineering and Computer Science, University of California, Irvine, California 92697, USA*

We present a general theory of exceptional points of degeneracy (EPD) in periodically time-variant systems that do not necessarily require the presence of loss or gain, and we show that even a single resonator with a time-periodic component may develop EPDs. An EPD is a special point in a system parameter space at which two or more eigenmodes coalesce in both their eigenvalues and eigenvectors into a single degenerate eigenmode. We demonstrate the conditions for EPDs to exist in time-periodic systems that are either lossless/gainless or with loss and/or gain and we show that a system with zero time-average loss/gain exhibits EPDs with purely real resonance frequencies, yet the resonator energy grows algebraically in time. We show the occurrence of EPDs in a single LC resonator while the introduced concept is general for any time-periodic system. These findings have significant importance in various electromagnetic/photonic systems and pave the way of applications in areas of sensors, amplifiers and modulators. A potential application of this time varying EPD is highlighted as a highly-sensitive sensor.

Frequency splitting phenomena at exceptional points of degeneracy (EPDs) is adopted to serve in sensing applications[1], [2]. Frequency splitting occurs at degenerate resonance frequencies where multiple eigenmodes of the system coalesce. Such degenerate resonance frequencies are extremely sensitive to small changes in the system which lead to a detectable shift in the system variables. This concept is used in modern sensing devices such as optical microcavities[2]–[4], bending curvature sensors[5] and optical gyroscopes[6], [7].

The splitting point of degenerate resonance frequencies varying a system parameter is referred to as an EPD and it emerges in a system when two or more eigenmodes coalesce in both their eigenvalues and *eigenvectors* into a single degenerate eigenmode. The concept of EPDs has received a surge of interest in recent years[8]–[16]. EPDs have been found in non-Hermitian parity-time (PT-) symmetric coupled systems, i.e., systems with balanced gain and loss[17], [9], [11], [14]. EPDs based on the concept of PT-symmetry have been investigated in coupled waveguides whose eigenmodes evolution is described in space[18], [8], [1], and also in coupled resonators where the eigenmodes evolution is described in time[10], [12], [15]. EPDs may also exist in lossless/gainless periodic waveguides which support multiple polarization eigenmodes that are periodically mixed and they usually occur at the transmission band edge[19]–[28], [13]. In essence, EPDs are obtained when the system matrix is similar to a matrix that contains a Jordan block[21], [29], [30]. At the EPD, the system eigenstate is represented in terms of generalized eigenvectors rather than the regular eigenvectors[20], [21]; which in turn leads to algebraic growth in the system eigenstates[21], [30]. There are different unique properties associated with the emergence of EPDs which lead to various potential applications such as enhancing the gain of active systems[31]–[35], [16], directivity in antennas[36], enhanced sensors[1], [2], [37], etc.

In this paper, we demonstrate the occurrence of EPDs in temporally periodic systems, in analogy to the EPDs found in spatially periodic structures[13], [19]–[22], [26], [28], [16]. Time variation was already considered in PT-symmetric systems with EPDs[38]–[42] where time variation was used to enhance some features in systems that already developed EPDs. However, in this paper we show that EPDs are induced in a single resonator directly due to periodic time-variation of the system itself. As an example, we demonstrate the concept using the simplest possible resonator, i.e., an LC resonator, though the formalism is general and applicable to any time periodic photonic or radio frequency system. Remarkably, we demonstrate that the occurrence of EPDs does not necessarily require the presence of gain or loss, hence EPDs are here demonstrated also in lossless/gainless linear time periodic (LTP) systems.

Time periodic variation is introduced in a system through any time-varying system parameter. Generally, an LTP system may comprise multiple components therefore we will assume that the state vector $\boldsymbol{\Psi}(t)$ describing this system is $N$-dimensional, i.e., $\boldsymbol{\Psi}(t) = [\Psi_1(t) \cdots \Psi_N(t)]^T$, where $T$ denotes the transpose operator. The temporal evolution of the state vector obeys the multidimensional first-order differential equation

$$\frac{d}{dt}\boldsymbol{\Psi}(t) = \underline{\mathbf{M}}(t)\boldsymbol{\Psi}(t) \qquad (1)$$

where $\underline{\mathbf{M}}(t)$ is the $N \times N$ time-variant system matrix. For LTP systems with period $T_m$, the state vector evolution from the time instant $t$ to $t+T_m$ is given by $\boldsymbol{\Psi}(t+T_m) = \underline{\boldsymbol{\Phi}}(t+T_m,t)\boldsymbol{\Psi}(t)$, where $\underline{\boldsymbol{\Phi}}(t+T_m,t)$ is the state *transition matrix*[43]. In the following, for simplicity and without loss of generality, we assume the matrix $\underline{\mathbf{M}}(t)$ is represented by a piecewise constant periodic function, hence we relate the transition matrix to the system matrices as $\underline{\boldsymbol{\Phi}} = \prod_{j=1}^{J} e^{\underline{\mathbf{M}}_j T_j}$, where $\underline{\mathbf{M}}_j$ is the system matrix in the $j^{\text{th}}$ interval $T_j$, and the system modulation frequency is $f_m = 1/T_m$ where $T_m = \sum_{j=1}^{J} T_j$.

Solutions of an LTP system in general satisfy $\mathbf{\Psi}(t+T_m) = e^{-i\omega T_m}\mathbf{\Psi}(t)$ which means that $\mathbf{\Psi}(t)$ can be represented in terms of the Floquet harmonics. Using the transition matrix to represent the time evolution of the state vector, we formulate the eigenvalue problem as

$$\underline{\mathbf{\Phi}}\mathbf{\Psi}(t) = \lambda \mathbf{\Psi}(t). \quad (2)$$

The eigenvalues $\lambda_n = \exp(-i\omega_n T_m)$ $n=1,\ldots,N$, with $\omega_n$ being the system eigenfrequencies, are obtained solving the characteristic polynomial $\det(\underline{\mathbf{\Phi}} - \lambda \underline{\mathbf{I}}) = 0$. When the transition matrix is diagonalizable we can write $\underline{\mathbf{\Phi}} = \underline{\mathbf{U}}\underline{\mathbf{\Lambda}}\underline{\mathbf{U}}^{-1}$, where $\underline{\mathbf{U}}$ is a non-singular similarity transformation matrix whose columns are the eigenvectors of $\underline{\mathbf{\Phi}}$ that are all independent. This analysis is valid unless an EPD emerges, at which the transition matrix $\underline{\mathbf{\Phi}}$ is non-diagonalizable and it is similar to a matrix $\underline{\mathbf{\Lambda}}_J$ that contains at least a non-trivial Jordan block with $P$ degenerate eigenvalues[30]. Therefore, at the EPD, the algebraic multiplicity $P$ of an eigenvalue $\lambda_e$ (and its correspondent eigenfrequency $\omega_e$) of (2) is higher than its geometrical multiplicity (the number of independent eigenvectors associated with that eigenvalue) because two or more eigenvectors coalesce. The similarity transformation at the EPD is written as $\underline{\mathbf{\Phi}} = \underline{\mathbf{V}}\underline{\mathbf{\Lambda}}_J\underline{\mathbf{V}}^{-1}$ where the columns of $\underline{\mathbf{V}}$ are composed of regular eigenvectors and generalized eigenvectors that are found through

$$(\underline{\mathbf{\Phi}} - \lambda_e \underline{\mathbf{I}})^P \mathbf{\Psi}_p(t) = \mathbf{0} \qquad p=1,2,\ldots,P \quad (3)$$

Here $\mathbf{\Psi}_p$ with $p>1$ are the generalized eigenvector and $P$ is the order of degeneracy (see Chap. 7 in[29]).

For the sake of simplicity, the following analysis and examples are focusing on second order EPDs that emerge in LTP systems, hence the transition matrix $\underline{\mathbf{\Phi}}$ has dimensions 2×2. For a system described by a real 2×2 transition matrix, the characteristic polynomial $\det(\underline{\mathbf{\Phi}} - \lambda \underline{\mathbf{I}}) = 0$ has real coefficients so that the eigenvalues $\lambda_1$ and $\lambda_2$ are either both real or a complex conjugate pair. Since $\det(\underline{\mathbf{\Phi}}) = \lambda_1 \lambda_2 = \prod_{j=1}^2 e^{\text{tr}(\mathbf{M}_j T_j)}$, where tr denotes the trace of the matrix, then either $\det(\underline{\mathbf{\Phi}}) = e^{2\,\text{Im}\{\omega_1\}T_m}$ when $\lambda_2 = \lambda_1^*$, where the symbol * denotes the complex conjugate operation, or $\det(\underline{\mathbf{\Phi}}) = e^{is\pi}e^{(\text{Im}\{\omega_1\} + \text{Im}\{\omega_2\})T_m}$ when $\lambda_1$ and $\lambda_2$ are both real, where $\omega_1$ and $\omega_2$ are the eigenfrequencies of the eigensystem (2) and $s$ is an integer. At the EPD the 2×2 transition matrix $\underline{\mathbf{\Phi}}$ is similar to a Jordan block with two degenerate eigenvalues that are associated with a regular eigenvector and a generalized eigenvector obtained from (3). On the other hand, the eigenvalues of any 2×2 matrix $\underline{\mathbf{\Phi}}$ are $\lambda_n = \text{tr}(\underline{\mathbf{\Phi}})/2 \pm \sqrt{[\text{tr}(\underline{\mathbf{\Phi}})/2]^2 - \det(\underline{\mathbf{\Phi}})}$ [44], where the upper and lower signs hold for $n=1,2$. The associated eigenvectors of the system described by a *non*-diagonal $\underline{\mathbf{\Phi}}$ are $\mathbf{\Psi}_n = [\phi_{12}, \lambda_n - \phi_{11}]^T$, $n=1,2$, where $\phi_{11}$ and $\phi_{12}$ are elements of the matrix $\underline{\mathbf{\Phi}}$ [43]. It is clear from the expression of the eigenvectors that degenerate eigenvalues lead to degenerate eigenvectors (i.e., identical eigenvectors) unless the matrix $\underline{\mathbf{\Phi}}$ is diagonal. Therefore, to guarantee the emergence of EPDs in a system described by a *non*-diagonal transition matrix $\underline{\mathbf{\Phi}}$, it is sufficient to have degenerate eigenvalues, i.e., it is sufficient to satisfy the condition

$$\text{tr}(\underline{\mathbf{\Phi}})/2 = \pm\sqrt{\det(\underline{\mathbf{\Phi}})}. \quad (4)$$

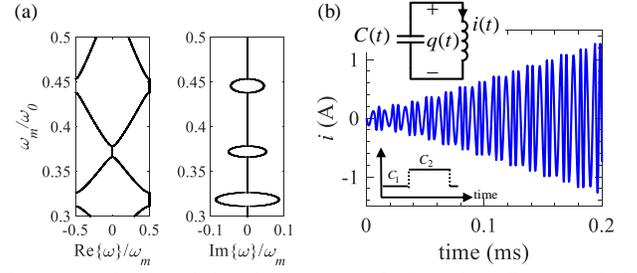

Fig. 1. (a) Real and imaginary parts of dispersion diagram of the eigenfrequencies versus the normalized angular modulation frequency. (b) The linear algebraic growth of the inductor current at an EPD for an LC resonator with time-varying capacitor.

The presented formulation is general and describes the occurrence of EPDs in any system described with the set of equations in (1). Without loss of generality and to provide a physical description of a 2×2 LTP system, we demonstrate all the above concepts by using very simple LC resonator examples.

Consider first the lossless LC resonator with time-periodic capacitance as depicted in Fig. 1(b); this example demonstrates the existence of EPDs in a simple LTP system made of a lossless and gainless resonator. We define the system state vector as $\mathbf{\Psi}(t) = [q(t)\ i(t)]^T$, where $q(t)$ is the instantaneous charge on the capacitor and $i(t)$ is the inductor current as shown in the circuit depicted in Fig. 1(b).

The time-variant capacitor is given by a two-level piecewise-constant time-periodic function: $C_1$ in the interval $0 \leq t < T_1$ and $C_2$ in $T_1 \leq t < T_m$. The time evolution of the state vector $\mathbf{\Psi}(t)$ is described by (1), where $\underline{\mathbf{M}}(t) = \underline{\mathbf{M}}_1$ for $0 \leq t < T_1$ and $\underline{\mathbf{M}}(t) = \underline{\mathbf{M}}_2$ for $T_1 \leq t < T_m$, with

$$\underline{\mathbf{M}}_j = \begin{bmatrix} 0 & -1 \\ 1/(L_0 C_j) & 0 \end{bmatrix}, \qquad j=1,2 \quad (5)$$

Fig. 1(a) illustrates the dispersion of the eigenfrequencies $\omega$ of $\underline{\mathbf{\Phi}}$, symmetrically located with respect to the center of the Brillouin zone (BZ), versus the normalized angular modulation frequency ($\omega_m/\omega_0$), where $\omega_0 = 1/\sqrt{L_0 C_0}$ and $C_0 = (C_1+C_2)/2$. Note that the system is periodic, so that for an eigenfrequency $\omega$ there correspond all the Floquet harmonics $\omega + s\omega_m$, where $s$ is an integer. For the system matrices (5) with real-valued elements, it can be shown that $\det(\underline{\boldsymbol{\Phi}}) = 1$ which implies either that $|\lambda_1| = |\lambda_2| = 1$ (hence $\mathrm{Im}\{\omega_1\} = \mathrm{Im}\{\omega_2\} = 0$) when $\lambda_2 = \lambda_1^*$, or that $\lambda_2 = 1/\lambda_1$ (hence $\mathrm{Im}\{\omega_1\} = -\mathrm{Im}\{\omega_2\}$) when both eigenvalues are real. This means that at the EPDs, the degenerate eigenvalue $\lambda_e = \pm 1$, therefore $\omega_e$ is purely real. This finding agrees with what is shown in the complex dispersion diagram depicted in Fig. 1(a). The parameters of the LC resonator are set as $L_0 = 10\mu\mathrm{H}$, $C_1 = 50\mathrm{nF}$, $C_2 = 150\mathrm{nF}$. The two $\omega$ solutions coalesce for some modulation angular frequencies $\omega_m$ and become exactly equal at the EPDs. In this particular example, EPDs occur at either the center of the BZ ($\mathrm{Re}\{\omega\}=0$) or at the edge of the BZ ($\mathrm{Re}\{\omega/\omega_m\} = \pm 0.5$). It is important to point out that for a lossless and gainless system the imaginary part of the dispersion diagram is symmetric with respect to the center of the BZ and it vanishes at the EPDs, i.e., EPDs occur at real valued frequencies.

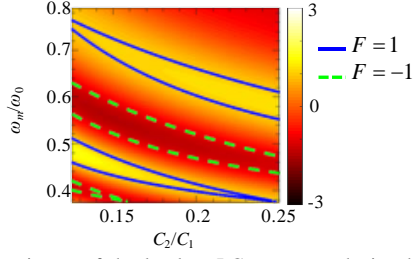

Fig. 2. The function $F$ of the lossless LC resonator depicted in Fig. 1(b) versus the capacitance $C_2$ and the normalized modulation angular frequency $\omega_m/\omega_0$. Solid and dashed contours on the color map represent points which satisfy the EPDs condition (6).

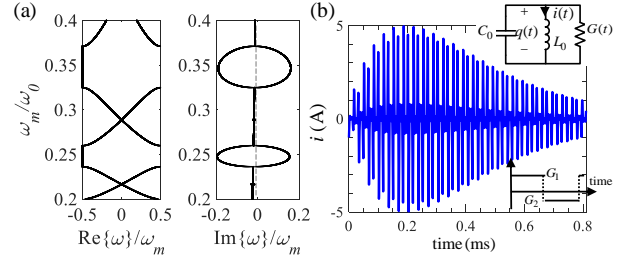

Fig. 3. (a) Dispersion diagram of the eigenfrequencies of the resonator depicted in (b), versus normalized modulation angular frequency. The vertical dashed line represents the $\mathrm{Im}\{\omega\}=0$ to point out that the imaginary part is negative. (b) Time evolution of the inductor current at an EPD for the time-periodic lossy resonator with positive time-average conductance as shown in the figure.

In Fig. 1(b) we show the time domain simulation of the state vector element $i(t)$ at one of the EPDs ($\omega_m = 0.438\omega_0$) assuming a capacitor charge initial condition of $q_c(0^-) = 50\,\mathrm{nC}$. It is clear from Fig. 1(b) that the capacitor current is growing linearly with time even in the absence of gain, which resembles one of the most important characteristics associated to the generalized eigenvector of a second order EPD. Note that, even though there is no gain element in the system, the time-periodic LC resonator is not isolated, and it receives energy from the time variation process.

The general condition (4) is sufficient for a system described by a non-diagonal 2×2 transition matrix to exhibit EPDs. In particular, for the lossless and gainless time-periodic LC resonator in Fig. 1 this sufficient condition reads

$$F = \cos(\Omega_1 T_1)\cos(\Omega_2 T_2)$$
$$-\frac{1}{2}\left(\sqrt{\frac{C_1}{C_2}} + \sqrt{\frac{C_2}{C_1}}\right)\sin(\Omega_1 T_1)\sin(\Omega_2 T_2) = \pm 1 \quad (6)$$

where $\Omega_j = (L_0 C_j)^{-1/2}$, with $j = 1, 2$, are the resonance frequencies of the LC resonator in the time intervals $T_1$ and $T_2$, respectively. Fig. 2 shows the function $F$ varying the capacitance $C_2$ ($C_1$ is constant) and the normalized angular modulation frequency ($\omega_m/\omega_0$). The solid blue and dashed green contours represent points on the colormap where an EPD exists: the blue contours represent points where $F = 1$ which are associated with EPDs at the center of the BZ, while the dashed green contours represent points where $F = -1$ that are associated with EPDs at the edge of the BZ.

As a second example we show that a system with time-periodic loss and/or gain also exhibits EPDs. To demonstrate the emergence of EPDs in this type of time-periodic systems, we use an RLC resonator with time-periodic piecewise constant conductance, as depicted in Fig. 3(b). Analogously to the previous example, we assume a two-level piecewise constant time-periodic conductance: $G_1$ for $0 \leq t < T_1$ and $G_2$ for $T_1 \leq t < T_m$. In general, $G_1$ and $G_2$ can be positive (representing loss in the system) or negative (representing gain). The state vector is defined as $\boldsymbol{\Psi}(t) = [q(t)\ i(t)]^T$, with $q(t)$ as the instantaneous charge on capacitor and $i(t)$ as the inductor current. The constant system matrices of the two-time intervals are given by

$$\underline{\mathbf{M}}_j = \begin{bmatrix} -G_j/C_0 & -1 \\ 1/(L_0 C_0) & 0 \end{bmatrix}, \qquad j = 1, 2 \ . \quad (7)$$

Such a system develops EPDs in two different scenarios; the first scenario is through using time-periodic conductance with non-zero time-average whereas the second one is through using time-periodic conductance with zero time-average, i.e.,

$$G_1 T_1 + G_2 T_2 = 0 \quad (8)$$

Not to mention, a system with positive time-average conductance has dominant loss and a system with negative time-average conductance has dominant gain. Moreover, a system with zero time-average has a time-periodic gain and loss balance.

*Non-zero time-average conductance*. Fig. 3(a) shows the complex dispersion diagram of a system with positive time-average conductance (i.e., $G_1 T_1 + G_2 T_2 > 0$). The parameters of the LC resonator are $L_0 = 10\mu\mathrm{H}$, $C_0 = 100\mathrm{nF}$, $G_1 = 101\mathrm{mS}$, $G_2 = -99\mathrm{mS}$, $T_1 = T_2 = T_m/2$. Fig. 3(b) shows the inductor current at an EPD with $\omega_m = 0.3711\omega_0$ and $q_c(0^-) = 50\,\mathrm{nC}$.

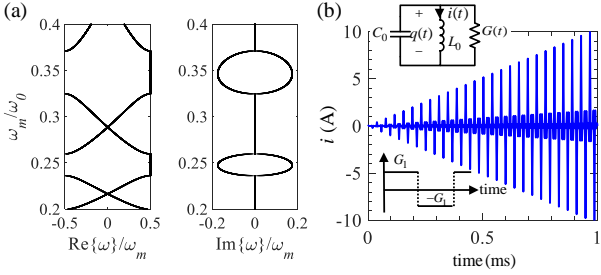

Fig. 4. (a) Dispersion diagram of the eigenfrequencies of the resonator depicted in (b), versus normalized modulation angular frequency. (b) Algebraic growth of the inductor current at an EPD of the time-periodic gain-loss balanced resonator, where the conductance is temporally periodic with zero time-average.

Note that for all the EPDs in such system, the imaginary part of the eigenfrequencies is negative which is a result of the dominant time-average loss in the system and that leads to an exponential decay of the system state vector besides the linear algebraic growth (visible at early times) due to the degeneracy.

Similar to the system with dominant time-periodic loss, a system with dominant time-periodic gain (i.e., $G_1T_1+G_2T_2 < 0$) will develop EPDs in its dispersion diagram. However, eigenfrequencies away from the bandgaps have a positive imaginary part which leads to an exponential growth of the state vector with time besides the linear growth at the EPD.

*Zero time-average conductance.* The determinant of the transition matrix is $\det(\underline{\mathbf{\Phi}}) = \prod_{j=1}^{2} e^{\mathrm{tr}(\mathbf{M}_j T_j)} = e^{-(G_1T_1+G_2T_2)/(2C_0)}$, therefore when (8) is satisfied, $\det(\underline{\mathbf{\Phi}}) = 1$ which leads to the same conclusion found for the lossless/gainless LC resonator discussed in relation to Fig. 1. This means that at the EPDs $\lambda_e = \pm 1$ and therefore $\omega_e$ is purely real.

Assuming $T_1 = \alpha T_m$ then $T_2 = (1-\alpha)T_m$ and assuming $G_1 > 0$, then to satisfy the time-average condition (8), the conductance $G_2 = -\alpha G_1/(1-\alpha)$.

The schematic of a time-periodic gain and loss balanced LC resonator is depicted in Fig. 4(b) where the time-variant conductance is assumed to satisfy (8). The dispersion diagram shown in Fig. 4(a) is based on parameters $L_0 = 10\mu H$, $C_0 = 100nF$, $G_1 = -G_2 = 100mS$ and $T_1 = T_2 = T_m/2$, and modulation frequency $\omega_m = 0.3248\omega_0$ with $\omega_0 = 1/\sqrt{L_0C_0}$. The time domain evolution of the inductor current $i(t)$, assuming an initial capacitance charge $q_c(0^-) = 50\,\mathrm{nC}$, is shown in Fig. 4(b). Since the system has a 2nd order EPD and a time-periodic gain and loss balance, we only observe a dominant linear growth in the current of the resonator at the EPD.

From (4) we find the sufficient condition for EPDs to emerge in a system with time-variant conductance to be

$$F = \cos(\Omega_1 T_1)\cos(\Omega_2 T_2)$$
$$- \frac{\omega_0^2}{\Omega_1 \Omega_2}\left[1 - \frac{G_1 G_2}{4G_0^2}\right]\sin(\Omega_1 T_1)\sin(\Omega_2 T_2) = \pm 1 \quad (9)$$

where $G_0 = \sqrt{C_0/L_0}$ and $\Omega_j = \omega_0\sqrt{1-G_j^2/(4G_0^2)}$, $j = 1,2$ are the real part of the complex resonance frequencies for the resonator in the time intervals $T_1$ and $T_2$, respectively. In Fig. 5(a) we show the function $F$ versus conductance $G_1$ and the normalized angular modulation frequency $\omega_m/\omega_0$ where we assume a zero time-average conductance with $\alpha = 0.6$. There exists a critical value of the conductance $G_1$, that is denoted by $G_c = 2G_0 \max\{1, (1-\alpha)/\alpha\}$, beyond which both $\Omega_1$ and $\Omega_2$ become purely imaginary and the system cannot exhibit EPDs. In Fig. 5(b), the solid blue contours ($F = 1$) and the dashed green contours ($F = -1$) represent points on the colormap where an EPD exists.

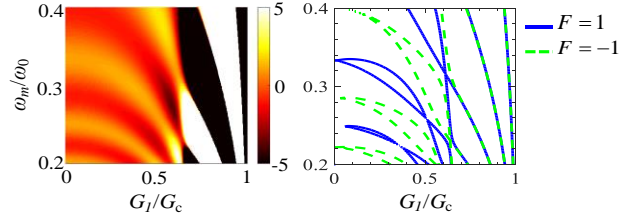

Fig. 5. (a) The function $F$ in terms of the conductance $G_1$ and the normalized modulation angular frequency $\omega_m/\omega_0$ for $\alpha = 0.6$. (b) Contours of points which satisfy the EPD condition in (9).

As discussed in the introduction, the eigenvalues at the EPDs are extremely sensitive to the perturbations of the system parameters. In general, introducing a perturbation $\varepsilon$ to any of the system parameters leads to a perturbed transition matrix $\underline{\mathbf{\Phi}}(\varepsilon)$ and perturbed eigenvalues $\lambda_p(\varepsilon)$, with $p = 1, 2, \ldots, P$ where $P$ is the order of the EPD. We represent $\lambda_p(\varepsilon)$ near the EPD eigenvalue $\lambda_e$ by a single convergent Puiseux series containing only powers of $\varepsilon^{1/P}$ where the Puiseux series coefficients are calculated using the explicit recursive formulas given in [45]. For a second order EPD ($P = 2$), we use a first order approximation of $\lambda_p(\varepsilon)$ as $\lambda_p(\varepsilon) \approx \lambda_e + (-1)^p \alpha_1 \sqrt{\varepsilon}$, where $\alpha_1 = \sqrt{-\partial f(\varepsilon,\lambda)/\partial \varepsilon}\Big|_{\varepsilon=0, \lambda=\lambda_e}$ and $f(\varepsilon,\lambda) = \det(\underline{\mathbf{\Phi}}(\varepsilon) - \lambda \underline{\mathbf{I}})$. As an illustrative example, consider an LC resonator with time-variant capacitance as described in Fig. 1, but with a lossy inductor with quality factor of 100, and assume $C_2$ is now perturbed from its nominal value as $(1 + \varepsilon) C_2$. When $C_2$ is not perturbed the system exhibits an EPD with an eigenvalue $\lambda_e = -0.9845$ for a modulation frequency $\omega_m = 0.438\omega_0$. In Fig. 6(a) we show the two perturbed eigenvalues $\lambda$ versus the perturbation $\varepsilon$ calculated from the exact eigenvalue problem (2) and by using the Puiseux series approximation. The most important thing to be noticed from Fig. 6 is that an extremely small perturbation in the capacitor $C_2$ will lead to a much larger change in the eigenvalues of the system. This property is actually one of the most exceptional physical properties associated to EPDs and it can be exploited in designing extremely sensitive sensors[8], [15]. The large perturbation of the eigenvalues in turn implies a sharp change in the complex resonance frequency of the LTP LC resonator as shown in Fig. 6(b). For positive but very small $\varepsilon$-perturbation the imaginary part of the complex resonance frequency shows a sharp change while its real part is kept constant. Furthermore, a very small negative $\varepsilon$-perturbation causes a rapid change of the real part of the resonance frequency. For example, a +0.1% perturbation in the dielectric permittivity of the capacitor $C_2$ (i.e., $\varepsilon = 0.001$) will change the imaginary part of the resonance frequency by $0.34\omega_0$; which implies 76% change in the quality factor of the resonator. Hence, this highly sensitive system can be employed to measure the dielectric permittivity (that can be changed by an environment parameter, like acidity or humidity, or a gas presence) with very high accuracy.

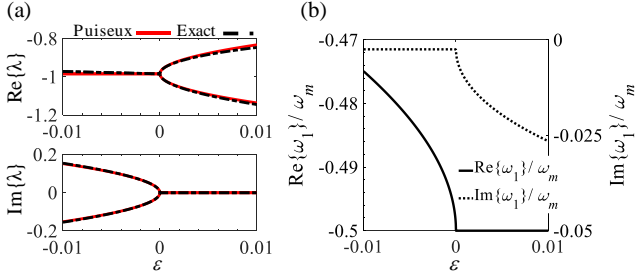

Fig. 6. (a) Variation of the eigenvalues away from the EPD of a lossy LC resonator as in Fig. 1(b) but with a lossy inductor, when the capacitor $C_2$ is perturbed. (b) The complex resonance frequency $\omega_1$ exhibits large variations even for very small perturbations $\varepsilon$.

## CONCLUSION

In conclusion, time-periodic systems show EPDs of different orders induced directly by time modulation of a parameter of the system. This does not need the presence of gain or loss in the system element. This concept has been demonstrated analytically and with examples with 2nd order EPDs. We have shown the analytic sufficient conditions for such EPDs to exist. We have demonstrated that in absence of any gain and loss circuit elements the resonance frequency of an EPD is purely real. Furthermore, this property holds also when gain and loss elements are introduced in the system assuming that the time-average of gain and loss is zero. Remarkably, at the EPDs the energy inside a time-periodic resonator grows algebraically with time even when the resonance frequency is purely real. Moreover, we have illustrated how such temporally induced EPDs lead to systems where the eigenvalue (and so the complex eigenfrequency) is exceptionally sensitive to a perturbation of the system. This may have potential application in conceiving highly-accurate sensing devices.

## ACKNOWLEDGEMENT

This material is based upon work supported by the Air Force Office of Scientific Research under award number FA9550-15-1-0280, and by the National Science Foundation under award ECCS-1711975. The authors acknowledge useful discussions with Prof. Franco Maddaleno, Polytechnic of Turin, and with Prof. Alexander Figotin, University of California, Irvine.

## APPENDIX: ANALITICAL TIME-DOMAIN SOLUTION OF THE EVOLUTION EQUATIONS

Analytical time-domain solution of the evolution equation in (2) can be obtained using the method explained in [43]. Using this method the state vector $\mathbf{\Psi}(t)$ at any time instance $t = mT_m + \chi$, with $\chi < T_m$ is

$$\mathbf{\Psi}(t) = \underline{\mathbf{\Phi}}(\chi,0)\underline{\mathbf{\Phi}}(T_m,0)^m \mathbf{\Psi}(0) \quad (A1)$$

where $\mathbf{\Psi}(0)$ is the initial condition of the state vector, and $\underline{\mathbf{\Phi}}(t_2,t_1)$ is the state transition matrix between the time instances $t_1$ and $t_2$

$$\underline{\mathbf{\Phi}}(t_2,t_1) = \underline{\mathbf{W}}(t_2)\underline{\mathbf{W}}^{-1}(t_1) \quad (A2)$$

here $\underline{\mathbf{W}}(t)$ is the Wronskian matrix of the solutions in the time interval between $t_1$ and $t_2$. As an example, we consider the LC resonator with time variant capacitor. In this example, we have two Wronskian matrices of the solutions associated to each time interval, $T_j$ is

$$\underline{\mathbf{W}}_j(t) = \begin{bmatrix} \cos\Omega_j t & \frac{1}{\Omega_j}\sin\Omega_j t \\ -\Omega_j \sin\Omega_j t & \cos\Omega_j t \end{bmatrix} \quad (A3)$$

where $\Omega_j$ with $j = 1, 2$ are the resonance frequencies of the LC resonator in the time intervals $T_1$ and $T_2$, respectively. Using the Wronskian matrix properties

$$\underline{\mathbf{W}}^{-1}(t) = \underline{\mathbf{W}}(-t) \quad (A4)$$

$$\underline{\mathbf{W}}(t_1)\underline{\mathbf{W}}(t_2) = \underline{\mathbf{W}}(t_1 + t_2) \quad (A5)$$

the transition matrix between time instances $t_1$ and $t_2$, when they are within the same time interval $T_1$ or $T_2$ is

$$\underline{\mathbf{\Phi}}(t_2,t_1) = \underline{\mathbf{W}}(t_2)\underline{\mathbf{W}}^{-1}(t_1) = \underline{\mathbf{W}}(t_2)\underline{\mathbf{W}}(-t_1) = \underline{\mathbf{W}}(t_2 - t_1). \quad (A6)$$

For time instances $t_1$ and $t_3$ within different time intervals the state transition matrix reads as

$$\begin{aligned}\underline{\mathbf{\Phi}}(t_3,t_1) &= \underline{\mathbf{W}}(t_3)\underline{\mathbf{W}}^{-1}(t_1) = \underline{\mathbf{W}}(t_3)\underline{\mathbf{W}}^{-1}(t_c)\underline{\mathbf{W}}(t_c)\underline{\mathbf{W}}^{-1}(t_1) \\ &= \underline{\mathbf{\Phi}}(t_3,t_c)\underline{\mathbf{\Phi}}(t_c,t_1)\end{aligned} \quad (A7)$$

where $t_c$ is the time instance which capacitor value changes. Therefore, the state transition matrices at different time instances $\chi$ for different pieces of the time variant capacitor reads as

if $\chi$ belongs to time interval $T_1$

$$\underline{\Phi}(\chi,0) = \begin{bmatrix} \cos\Omega_1\chi & \frac{1}{\Omega_1}\sin\Omega_1\chi \\ -\Omega_1\sin\Omega_1\chi & \cos\Omega_1\chi \end{bmatrix} \quad (A8)$$

or if $\chi$ belongs to time interval $T_2$

$$\underline{\Phi}(\chi,0) = \underline{\Phi}(\chi,T_1)\underline{\Phi}(T_1,0)$$
$$= \begin{bmatrix} \cos\Omega_2(\chi-T_1) & \frac{1}{\Omega_2}\sin\Omega_2(\chi-T_1) \\ -\Omega_2\sin\Omega_2(\chi-T_1) & \cos\Omega_2(\chi-T_1) \end{bmatrix} \quad (A9)$$
$$\begin{bmatrix} \cos\Omega_1 T_1 & \frac{1}{\Omega_1}\sin\Omega_1 T_1 \\ -\Omega_1\sin\Omega_1 T_1 & \cos\Omega_1 T_1 \end{bmatrix}$$

Using state transition matrices in (A8) and (A9) we can find the solution to the evolution equation in (2) knowing the initial state of the system.